%% file: main.tex
\documentclass[acmlarge]{acmart}

\usepackage{booktabs} 
\usepackage[ruled]{algorithm2e} 

\SetAlFnt{\small}
\SetAlCapFnt{\small}
\SetAlCapNameFnt{\small}
\SetAlCapHSkip{0pt}
\IncMargin{-\parindent}

\acmJournal{IMWUT}
\acmVolume{00}
\acmNumber{00}
\acmArticle{00}
\acmYear{2019}
\acmMonth{00}

\usepackage{amsmath}
\usepackage{amsfonts}
\usepackage{bm}
\usepackage{makecell}
\usepackage{multirow}

\usepackage[textsize=tiny]{todonotes}

\setcopyright{usgovmixed}

\begin{document}
	\title{Towards Reliable, Automated General Movement Assessment for Perinatal Stroke Screening in Infants Using Wearable Accelerometers}	
	\author{Yan Gao*}
	\affiliation{%
		\institution{Open Lab, School of Computing, Newcastle University}
		\city{Newcastle upon Tyne}
		\country{UK}
	}
	\email{y.gao47@newcastle.ac.uk}
	
	\author{Yang Long*}
	\thanks{* these authors contribute equally with joint first authorship}
	\affiliation{%
		\institution{Open Lab, School of Computing, Newcastle University}
		\city{Newcastle upon Tyne}
		\country{UK}
	}
	\email{yang.long@ieee.org}
	
	\author{Yu Guan}
	\affiliation{%
		\institution{Open Lab, School of Computing, Newcastle University}
		\city{Newcastle upon Tyne}
		\country{UK}
	}
	\email{yu.guan@newcastle.ac.uk}
	
	\author{Anna Basu}
	\affiliation{%
		\institution{Institute of Neuroscience, Newcastle University, and Department of Paediatric Neurology, Great North Childrens Hospital}
		\city{Newcastle upon Tyne}
		\country{UK}
	}
	\email{anna.basu@newcastle.ac.uk}
	
	\author{Jessica Baggaley} 
	\affiliation{%
		\institution{Institute of Neuroscience, Newcastle University}
		\city{Newcastle upon Tyne}
		\country{UK}
	}
	\email{jess.baggaley2@newcastle.ac.uk}
	
	\author{Thomas Ploetz}
	\affiliation{
		\institution{School of Interactive Computing, Georgia Institute of Technology}
		\city{Atlanta}
		\country{USA}}
	\email{thomas.ploetz@gatech.edu}

\begin{abstract}
Perinatal stroke (PS) is a serious condition that, if undetected and thus untreated, often leads to life-long disability, in particular Cerebral Palsy (CP). 
In clinical settings, Prechtl's General Movement Assessment (GMA) can be used to classify infant movements using a Gestalt approach, identifying infants at high risk of developing PS. 
Training and maintenance of assessment skills are essential and expensive for the correct use of GMA, yet many practitioners lack these skills, preventing larger-scale screening and leading to significant risks of missing opportunities for early detection and intervention for affected infants. 
We present an automated approach to GMA, based on body-worn accelerometers and a novel sensor data analysis method--Discriminative Pattern Discovery (DPD)--that is designed to cope with scenarios where only coarse annotations of data are available for model training. 
We demonstrate the effectiveness of our approach in a study with 34 newborns (21 typically developing infants and 13 PS infants with abnormal movements). 
Our method is able to correctly recognise the trials with abnormal movements with at least the accuracy that is required by newly trained human annotators (75\%), which is encouraging towards our ultimate goal of an automated PS screening system that can be used population-wide.

\end{abstract}

	%
	%

	%
	%
	%


\maketitle

\renewcommand{\shortauthors}{}
\title[]{Towards Reliable, Automated General Movement Assessment for Perinatal Stroke Screening in Infants Using Wearable Accelerometers}

\input{intro}

\input{background}

\input{methodology}
\input{algorithm}

\input{experiments}
\input{conclusion}

\section*{Acknowledgments}
This work was supported jointly by Medical Research Council (MRC, UK) Innovation Fellowship (MR/S003916/1), Engineering and Physical Sciences Research Council (EPSRC, UK) Project DERC: Digital Economy Research Centre (EP/M023001/1), and National Institute of Health Research (NIHR, UK) Career Development Fellowship (CDF-2013-06-001)(AB). The views expressed are those of the authors and not necessarily those of the NHS, the NIHR, or the Department of Health and Social Care (DHSC, UK).

\bibliographystyle{ACM-Reference-Format}
\bibliography{main}

\end{document}

%% file: intro.tex

\section{Introduction}
Perinatal stroke (PS), i.e., a stroke occurring before or around the time of birth of an infant, is a serious concern because--if undetected and thus untreated--it can lead to conditions such as Cerebral Palsy (CP) with negative impact on quality of life and indenpendence.
CP is a term used to describe a group of disorders with lifelong adverse effects on movement and posture due to damage incurred to the developing brain.
It is the most common motor disorder in childhood, affecting around 2 in 1,000 children \cite{Johnson:2002vo}. 
A common form is unilateral CP, with weakness and stiffness affecting one side of the body. 
Unilateral CP is often caused by PS. Around 1 in 3,500 infants sustains PS, and between 10--50\% of infants with PS develop CP \cite{Nelson:2007gz}.

Early therapy intervention in unilateral PS is under investigation \cite{Basu:2017be,Basu:2018fg} and may improve motor outcomes  \cite{2014_busu}.
However, early intervention requires early detection of affected infants. 
This is challenging, as PS does not present with immediate signs of unilateral weakness, in contrast to stroke in adults;
clinical features at the time of stroke are less specific and in some cases not detectable, with later emergence of motor problems over several months.
Furthermore, there is no routine screening program for PS. 
Cranial Magnetic Resonance Imaging (MRI) could detect PS but is costly (e.g., \pounds 1,000 per scan) \cite{Edwards:2018kb}.
It is performed if indicated based on the clinical condition, i.e., if there is a high level of suspicion of neurological abnormality. If the infant is extremely preterm, cranial ultrasound screening is usually undertaken but may miss cases of PS \cite{Cowan:2005fd}.

Another option for screening and early detection of infants at risk of motor disorders such as CP is a standardised, manual visual observation procedure according to Prechtl's General Movements Assessment (GMA) \cite{2005_GMA}. It is used both clinically and in research to monitor infant movements in the first several months of life with high predictive validity \cite{Kwong:2018ij}. The GMA relies on the recognition and distinction of specific aspects of the infants' spontaneous movement repertoire during the first months of life. However, it is not a diagnostic test --- it highlights infants in need of further investigation and likely intervention.
The GMA requires the infant to be placed in the supine position for around 5-10 minutes in a quiet, alert state whilst being video recorded. 
The video is later scored by a trained professional (see Sec.\ \ref{sec:background:gma} for more details).
Reliable use of the GMA requires attendance on a specific training course, followed by extensive practice to obtain and maintain the skills required for Gestalt detection of abnormal infant movements.
Many practitioners do not currently have these skills.
Furthermore, the assessment is inherently subjective, and observer performance may be affected by fatigue, inter-subject variability, etc.


With the proliferation of wearable and pervasive sensing, combined with breakthroughs in machine learning based sensor data analysis, we hypothesise that automated GMA-based screening of infants can be realised.
Our goal is to develop methods that enable population-wide, minimal effort yet accurate and objective assessments of every newborn that will lead to early detection of potential motor abnormalities. 
Such an automated screening procedure will not lead to fewer cases of PS but to earlier detection.
The earlier PS can be detected, the earlier can dedicated treatments be administered to improve outcome.
The focus of our research agenda is to develop simple, high-fidelity assessments, which are the basis for accurate and objective diagnosis.

In this paper we lay the foundation for the aforementioned agenda.
Tri-axial accelerometers are attached to each limb of typically developing infants and those with PS to record their movements.
Through a machine-learning based analysis pipeline these sensor data are then automatically analysed with the objective of identifying infants who show signs of abnormal movements (AM) that may be indicative for PS.
Sensor data are of high temporal resolution, which is of importance for the detection of even subtle indicators of PS (according to GMA).
We have collected a dataset from 34 children (21 typically developing, and 13 with PS) who were observed as part of a research study.
Data were collected in multiple trials each, at monthly intervals and each infant provided at least one trial between the age of 1 and 6 months. 
Video recordings of the trials were reviewed separately by a trained professional blinded to the clinical condition, and classified (according to GMA criteria) as normal or abnormal.

Strictly speaking the automated sensor data analysis resembles a human activity recognition (HAR) problem, which has been widely studied in the ubiquitous computing community including many applications in health and wellbeing (e.g., \cite{Avci2010-ARU,2012_Plotz_SB,Kranz:2013_coach,2015_Hammerla_PD,Hoey2011-RSA,Ploetz2010a}.
A specific challenge that is linked to GMA-related movement assessment in infants (but not uncommon in the wider health-related HAR domain, e.g.\ \cite{2015_Hammerla_PD}) lies in the sparse annotation of the sensor data that can be used for modelling, which prevents the use of standard activity recognition pipelines such as \cite{2014_Bulling_tutorial}, or even more contemporary deep learning based methods \cite{2015_Yang_1dCNN_HAR,2016_roggen_ConvLSTM,2016_Hammerla_IJCAI,2017_Guan_LSTMensemble,Zeng:2014tk} that are notorious for relying on large amounts of annotated sample data.
Only one label is available per trial (assessment session) that in itself typically has a duration of several minutes: the infant shows abnormal movements or not.
In real-world clinical scenarios it is simply not feasible to provide sample-precise annotation that would allow to model dedicated events that are linked to abnormal movements.
To address this sample/ annotation imbalance problem we have developed a novel analysis framework that explicitly focuses on sparse labeling.
We present a Discriminative Pattern Discovery (DPD) framework that can suppress less informative segments for detecting behaviour patterns that corresponding to abnormal movements, which enables robust modelling that leads to correct and reliable automated assessments.
Based on an experimental evaluation using the recorded dataset we demonstrate that these automated assessments are of comparable accuracy to those provided by human experts, which is encouraging for our overall agenda of developing screening methods that can be used at population scale.
Given that wearable accelerometers are now very inexpensive and can thus be considered commodity devices, and the minimal effort needed for data recording, the proposed solution can be considered the first milestone towards our goal of generalised screening.


The contributions of this paper can be summarised as follows:
\begin{description}
	\item[Approach:] We developed a method for automated detection of abnormal movements (AM) in infants. 
	Through employing inexpensive, wearable accelerometers the assessment method is straightforward to integrate into everyday practice with minimal effort, which has the potential to significantly alleviate the cost of medical-expert based GMA.
	By employing machine learning based sensor data analysis methods, confounding factors of manual GMA assessments can be overcome, including skill and experience (or lack thereof) of a human assessor, fatigue, inter-subject variability, subjectivity, etc. 
	\item [Dataset:] We have collected a considerable dataset of 34 infants (21 typically developing, 13 with perinatal stroke) wearing the lightweight accelerometers on their limbs during multiple assessment sessions (trials). 
	This dataset includes realistic trial-wise expert annotations for model training and evaluation.
	\item [Analysis Method:] In response to the challenges of the data recording and especially annotation procedure we have developed a Discriminative Pattern Discovery (DPD) framework for modeling that effectively tackles the trial-wise classification tasks as a weak labelling problem. 
	We employ a novel kernel-based algorithm to improve model generalisation.
	\item [Model Performance:] Based on the recorded dataset we demonstrate that our automated analysis method can surpass the pass mark (75\% in accuracy) for the GMA exam required for a human to practise this method.
\end{description}

%% file: background.tex
\section{Background and Motivation}
\label{sec:background}
\subsection{Clinical Routine for Diagnosing Perinatal Stroke}
\label{sec:background:gma}
Around 10-50\% of infants who suffer from PS develop CP \cite{Nelson:2007gz}, due to damage to those parts of the brain that control movement, balance, and posture.
Through observation of spontaneous movements of an infant, and through neurological examinations supported by cranial imaging, the clinical diagnosis of CP can be made.
Although cranial imaging was used to provide a definitive gold standard by which to classify the infants as having had PS or not, definitive imaging with MRI is costly (around \pounds 1,000 per scan for infants in the UK), and requires sedation or general anaesthetic in some cases, which has an associated risk \cite{Salerno:2018dz}.
Therefore, MRI-based brain imaging of infants is not a routine screening procedure, even for preterm infants, though it is used as diagnostic evaluation of symptomatic infants.

On the other hand, movement monitoring for infants is widely applied in predicting impairments in neuro-motor development, and various qualitative methods have been proposed to assess the quality of motion patterns for infants \cite{2005_GMA}.
A more often used  approach for identifying infants at high risk is the use of Prechtl's General Movements Assessment (GMA) \cite{Einspieler:2005hq}.
Specifically, the term "General Movements" describes infant movements in the quiet alert state based on a Gestalt perception of normal versus abnormal movements that naturally encompasses factors such as the complexity, fluency and variation of the movements, and from around 3 months, the emergence of "fidgety" movements in the typically developing infant  \cite{2000_GM} \cite{1990_GM}.
GMA can be done within several minutes and it is purely observational.
However, it cannot classify infants as having PS or not: it can merely describe infant movements as being abnormal or normal.
GMA provides prognostic information if used correctly and thus infants identified on a clinical basis as having abnormal GMA would be followed up and investigated  (including brain imaging). 
The main difficulty with this assessment is that it requires extensive observer training and practice to retain the skills required for Gestalt detection of abnormal movements --- many practitioners do not have these skills. 

The clinical classification process according to GMA is age-dependent, because the characteristics of spontaneous infant movements evolve with age. 
Furthermore, the clinical classification of General Movements is based on the Gestalt impression of the quality of movement over the whole segment of the video recording viewed, of a supine infant in the quiet, alert state, paying attention to complexity, fluency and variation. 
GMs are scored as normal or abnormal based on this impression. 
GMs in the first two months of life (corrected for prematurity, i.e., from the expected date of delivery onwards) are predominantly "writhing" in nature, with relatively slow, fluent, elegant, complex, variable rotational movements. 
The three sub-classifications of types of abnormal GMs in this age group are "poor repertoire" GMs (lack of complexity and variety in the movement repertoire), "cramped-synchronised" GMs (with periods of synchronous whole body muscle contraction and relaxation) and "chaotic GMs" (extreme lack of movement fluency as well as large amplitude of movements). 
From age 6 weeks onwards, "fidgety" movements may begin to emerge; they peak in their frequency of occurrence at around 12--16 weeks and are typically observed until around 20 weeks though may persist a few weeks longer. 
Fidgety movements are smaller in amplitude, and higher in speed, than writhing movements. 
It is abnormal for fidgety movements to be absent, or to be present but very large amplitude and excessively fast and jerky. 
Thus, for infants aged 3 to 5 months (corrected for prematurity), the two sub-classifications of abnormal types of GMs are "absent fidgety movements" and "abnormal fidgety movements". 
For infants aged 6 months, voluntary movements are normally predominant; movements for infants aged 6 months were therefore classified as normal or abnormal by an experienced observer based on whether appropriate voluntary movements were observed.
Table \ref{tab:GMA} summarizes the types of abnormal General Movements as they can be observed at different age groups.

\begin{table}[t]
	\centering
	\caption{\label{tab:GMA} Types of abnormal General Movements as they are assessed in Prechtl's General Movement Assessment.}
	\begin{tabular}{llp{7cm}}
	\Xhline{1pt}
	Age group & Types of abnormal GM & Explanation \\
	\hline
	0 -- 2m & Poor repertoire & Lack of complexity and variety in the movement repertoire \\
	& Cramped-synchronised & Periods of synchronous whole body muscle contraction and relaxation \\
	& Chaotic & Extreme lack of movement fluency as well as large amplitude of movements\\
	\hline
	3 -- 5m & Absent fidgety movements & No fidgety movements seen (NB in clinical practice this would be confirmed by repeat examination later in the same age bracket) \\
	& Abnormal fidgety movements & Fidgety movements present but very large amplitude and excessively fast and jerky
	\end{tabular}
\end{table}


\subsection{Movement Assessment with Wearables}
\label{sec:background:har}
Recently, wearable technologies have been used for automated capturing and subsequent analysis of spontaneous movements of infants with a view to reproducing the GM classification without the need for a clinical observer \cite{Marcroft:2015tn}. 
In \cite{2012_Dana}, a clinical tool based on accelerometers was used to assess motion patterns for preterm infants and it was demonstrated that using accelerometery data was a reliable way to evaluate the characteristics of movement disorders \cite{2010_Heinze}.
Singh et al.\ developed a system that leverages accelerometers on detecting features related to CP through analysing abnormal movements in premature infants \cite{2010_ISWC}. 
Fan et al.\ developed a Markov model-based technique that can recognise gestures from accelerometers.
They demonstrated that by treating instantaneous machine learning classification values as observations and explicitly modelling duration, the recognition rate of abnormal movements can be improved \cite{2012_Fan}. 

From the perspective of ubiquitous computing, the aforementioned techniques fall in the field of human activity recognition (HAR), which plays a major role in computational behaviour analysis, with applications in, for example, daily life monitoring \cite{2011_Plotz_RBM}, medical diagnosis \cite{2010_FOG} \cite{2012_Plotz_SB}, sports tracking and coaching \cite{Kranz:2013_coach}, skill assessment \cite{2015_Khan_skill_assessment}\cite{2013_Ladha_climbAX}, health and welling being assessment \cite{2015_Hammerla_PD},  to name but a few.
In general, the goal of HAR is to automatically recognize what a person is doing, and when.
Over the years a multitude of methods have been developed, which is not surprising given that activity recognition constitutes one of the main pillars of ubiquitous computing contributing to automated context inference \cite{Abowd:2012vt}.
It is beyond the scope of this paper to give a detailed overview of activity recognition techniques, much of which can now even be considered somewhat common sense in the field (for excellent surveys on the field cf., for example, \cite{2014_Bulling_tutorial,6208895,Avci2010-ARU,HAR_DL2018}).

The predominant analysis approach is based on transforming raw sensor data into a sequence of isolated analysis frames by employing a sliding-window procedure \cite{2014_Bulling_tutorial}.
This sliding window procedure needs to be optimized very carefully to determine appropriate length of the analysis frames and overlap between subsequent frames.
Typically these parameters are optimized heuristically and globally for all activities of interest (even though recently variations with optimised window lengths for each activity of interest have been developed \cite{Li:2018ia}).
Subsequently, for every analysis frame features are extracted that are fed into classification backends that automatically decide upon the labels, that is class (activity) associations, of the portions of sensor data that are captured by the individual frames \cite{Figo2010-PTF,2013_Hammerla_ECDF,Kwon:ig}.
This standard activity recognition chain follows the supervised modeling approach and fine grained ground truth annotation is required for model training and optimisation: every sensor reading (sample) needs to be annotated, which is typically realized through manual observation and annotation on the recording timeline.
Implicitly this annotation is translated into a sample-wise labelling but it requires a human observer to pay very close attention to the annotation process itself. 
For real-world deployment scenarios such time-consuming annotation procedures--especially for the sake of collecting annotated sample data for method development--are often unfeasible as they interfere with the primary task such as providing care.
Instead more coarse annotation schemes are employed such as providing assessments on a session or trial level, as pursued in our application scenario.
On a technical level this trial-wise annotation translates into 
a weak supervision problem with few labels only for many sensor readings, which introduces ambiguity that rules out standard supervised model training procedures.
Our analysis method effectively addresses this weak supervision problem through automatically discovering the subtle discriminative patterns within the recordings of complete analysis sessions that can differentiate abnormal movements (AM) from typical developing (TD) infants. 

\subsection{Model Generalisation}
\label{sec:background:generalisation}
Our sensor data analysis approach is based on machine learning techniques, for which we derive the parameters of a probabilistic model from sample data. 
This model training aims at generalisation such that the model is not only capable of analysing the training data but also performs well on unseen data, which resembles the intended application case.
This generalisation 
is one of the most pressing concerns in the wider machine learning research field.
It is expected that a trained model can be applied to new situations or environments, and  
existing approaches roughly fall into three categories: 
\textit{i)} problem-based;
\textit{ii)} model-based; and 
\textit{iii)} sample-based strategies. 
The latter aims to discover more information from different training sources or different views, such as in ensemble learning \cite{ensemble, 2017_Guan_LSTMensemble, 2015_ensembleGait}. 
Problem-based approaches consider special assumptions of different tasks and develop specialised variants for generalisation. 
For example, transfer learning assumes one or more additional modalities that are drawn from the same joint distribution to the problem domain. 
Model-based generalisation focuses on generic models of better representation or classifiers, such as dictionary learning, deep learning, and manifold learning.
This paper contributes to the family of problem-based modeling, in particular Multiple Instance Learning \cite{1997_MI,2004_GMIL}. 
Our key extension is to relax the strict assumption by a soft proportion according to our problem. 
Furthermore, we propose an elegant and straightforward implementation using kernel embedding \cite{2001_PDF}. 


\subsection{Towards Automated General Movement Assessment for Perinatal Stroke Screening}
\label{sec:background:vision}
The overarching goal of our research is the development and validation of automated ways for population-wide screening infants for abnormal movements due to PS and other early acquired brain injuries.
The earlier these can be discovered, the earlier specific care and treatment programs can be started. 
Screening could be undertaken on the delivery suit and as part of routine visit by health visitors, and midwives, or even by the parents at home if the recording procedure is as straightforward as possible.
Movement data will then be analysed automatically and potential findings communicated to both caregivers and paediatricians.
During wellbeing checkups the data can then be discussed and further actions be planned. 
The use of commodity sensing hardware lowers the burden of deployment.
We use miniaturised inertial measurement units that are straightforward to use by integrating the sensing platforms into pouches that can easily be attached to the limbs, not hindering the infant's natural movements.
The more straightforward the data recording procedure, the higher the quality of the collected data, which increases the chances for successfully and automatically recognising abnormal movements.

%% file: methodology.tex
\section{Methodology}
\label{sec:methodology}
\subsection{Data Acquisition}
\label{sec:methodology:data_acquisition}

\noindent
\paragraph{Participants} 
We recruited 21 Typically Developing (TD) infants and 13 infants with PS. 
Ethical approval was obtained prior to the study (West of Scotland Research Ethics Committee, reference number 15/WS/0129), and all necessary research governance processes were followed.
Infants were excluded if they: 
\textit{i)} had any additional significant medical diagnoses, such as severe visual impairment, which could render outcomes uninterpretable. 
\textit{ii)} had radiological evidence of significant bilateral intra cerebral pathology or that only the occipital, prefrontal or temporal lobes (i.e., non-motor areas) were affected; 
\textit{iii)} had ongoing involvement in another study which would likely interfere with this study. 
Descriptive statistics of the involved participants' assessments can be found in Table \ref{tab_participants}.
\begin{table}[]
	\centering
	\caption{Statistics of Assessments by Age, Diagnosis and Gestation (\#:number of trials)}
	\label{tab_participants}
	\begin{tabular}{ccccccc}
		\Xhline{1pt}
		\multirow{2}{*}{Month} & \multicolumn{3}{c}{Typical Developing} & \multicolumn{3}{c}{Perinatal Stroke} \\\cline{2-7}
		& \#Term       & \#Preterm      & \#Total      & \#Term      & \#Preterm      & \#Total     \\\hline
		1                      & 6          & 3            & 9          & 2         & 6            & 8         \\
		2                      & 11         & 4            & 15         & 5         & 6            & 11        \\
		3                      & 12         & 4            & 16         & 5         & 7            & 12        \\
		4                      & 14         & 4            & 18         & 5         & 5            & 10        \\
		5                      & 14         & 3            & 17         & 5         & 4            & 9         \\
		6                      & 16         & 4            & 20         & 5         & 5            & 10     \\\hline
		Total & 73 & 22 & 95 & 27 & 33 & 60 \\\hline
	\end{tabular}
\end{table}

\begin{figure}
	\centering
	\includegraphics[width=0.5\textwidth]{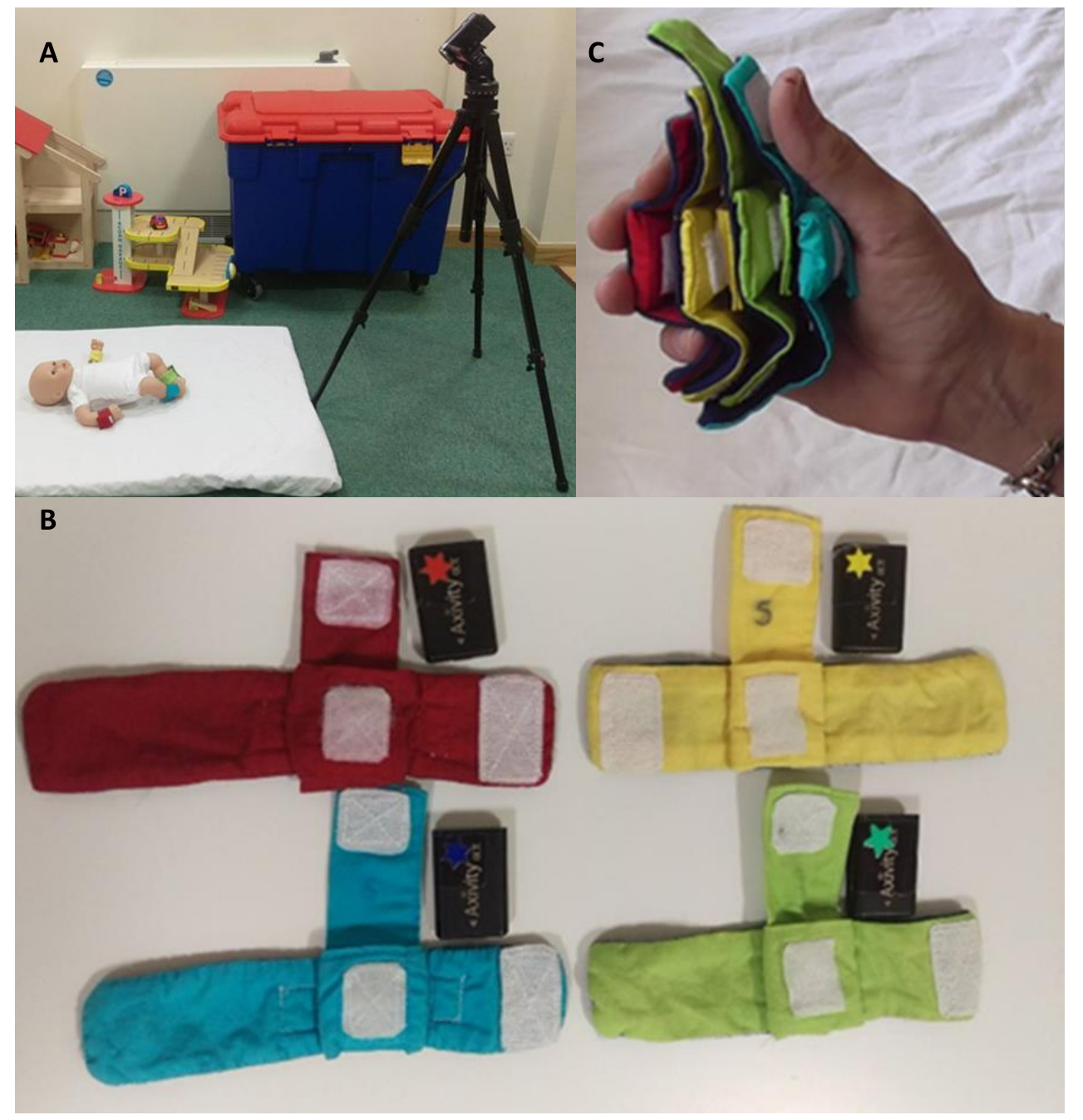}
	\caption{(A) During the ten minute supine recordings, infants were placed onto a mat with their navel around 91cm from the centre base of the camera.
	(B) Colour coded IMUs were placed into matching colour coded wrist and ankle straps. 
	All IMUs were face up with the Axivity logo showing and the micro-USB port arrow facing into the pouch. 
	The pouch has been specifically designed to tightly fit an IMU, when sealed in a plastic bag, to reduce coincidental movement. 
	(C) The straps were held together with the pouches on top of one another and shaken on camera to create a visual pattern that was used to synchronise the data. \label{fig_dataCollect}}
	\vspace*{-1em}
\end{figure}

\noindent
\paragraph{Data Collection} 
All participants underwent a ten-minute recording whilst lying supine at monthly intervals of their birthday from term age, or term equivalent age for preterms, for six months. 
To record movement during the supine recording, infants wore ankle and wrist straps, specifically designed for use with Axivity WAX9 inertial movement unit (IMUs \cite{wax9} which includes accelerometer, gyroscope and magnetometer) recordings. 
In this work, only tri-axial accelerometer data were used. 
The straps were made of four brightly coloured cottons (yellow, blue, green and red, as shown in Fig.\  \ref{fig_dataCollect}) with a sealable IMU pouch and an individual piece of Velcro hooks sewn onto one end of the strap. The inside of the strap was made of pile felt which would stick to the Velcro hooks therefore creating a secure, adjustable comfortable fit for each infant. 

The IMUs are light weight (approx. 10g) with recorded acceleration at 100Hz in three axes (x, y and z). 
The WAX9 devices were switched on and sealed into waterproof bags prior to being inserted into the pocket of the straps, with the Axivity logo arrow face up and pointing into the strap's pouch. 
To synchronise the IMUs, short impulse forces, created by shaking all the IMUs at once, were used to create data markers \cite{Ploetz:2012ts}. 
At the end of the recordings the IMUs were subjected to a second set of short impulse forces at the same time in front of the video camera to provide a second set of data markers.
The whole process was also recorded by a camera for annotation purposes. 
By following this protocol we inserted markers into both the sensor recordings and the video footage, which  enabled time-alignments between the data streams.
The data collection environment is illustrated in Fig. \ref{fig_dataCollect}.

The data recording procedure is straightforward because it does not interfere with normal infant handling routine.
The sensors can easily be slipped into the pouches.
The tight fit and the clear indications of orientation alignment greatly reduces the risk of misalignment or inserting the sensors with the wrong orientation.
The pouches are integrated into colour-coded straps that easily fit the limbs of newborns. 
With their fully adjustable fit and range of sizes, they are easy to put on.
It takes only a few seconds to attach the sensor to a limb, a routine that can easily be integrated into general handling of the infant, e.g., during play.

We also recorded some meta information for this dataset. 
For example, 'head position' was recorded based on the position of the infant's head as 
either `Left', `Right' or `Midline'. 
`Left' and `Right' were determined as the infant having turning their head over 30\textdegree\ away from the midline in either direction. 
In order to constrain the requirements of our approach to what is feasible in real-life application scenarios we did not use the head position information, even though it may provide important context. 
The other meta information we recorded is the quality of the trials. 
Unqualified trials were excluded for this study, such as 
when the infant had rolled their front, or the infant was crying/ sneezing/ sick/ coughing etc. 
As a result, we obtained a number of continuous trials each 4-10 minutes in length. 
A professional trained in GMA observed each video and classified infant movements as normal (TD) or abnormal (AM). 
Age information were also included, as shown in Table. \ref{tab_participants}.

\subsection{Data Pre-processing}
\label{sec:methodology:preprocessing}
After removing some unqualified/ corrupted trials, we had 161 validated continuous trials (some longer trials were broken into shorter ones). 
Each trial contained four sets of synchronised sequences that corresponded to the four accelerometers fixed on the infant's four limbs. 
Each accelerometer dataset (collected from a limb) consisted of three axis data (x, y and z) and their Signal Vector Magnitude as a complementary dimension, i.e., 4 channels (per limb). 
The trial length was roughly 4-10 minutes, i.e., 24k-60k of samples. 
Each trial was labeled indicating whether the data came from a typical developing (TD) infant or an infant with abnormal movements (AM). 
In total, there were 64 positive trials (AM) and 97 negative trials (TD).


For data processing, we employed the sliding window approach without any window/frame overlapping.
We further concatenated each 1s window (i.e., 100 samples) of $16$ channels raw data (i.e., from 4 limbs) into a $16\times 100$-dimensional vector. 
As a result, each trial was segmented into a number of windows, while each window corresponded to 1s movement records. 
We aimed to identify windows of the abnormal movements (wAM), and most importantly also to make an overall decision (i.e., AM/ TD) for the whole trial, which corresponded to the Gestalt nature of GMA.
To further improve the efficiency and remove redundancy, we conduct PCA dimension reduction to project each window of raw signals into a low dimensional feature space for further processing.

\subsection{Challenges}
\label{sec:methodology:challenges}
\noindent
\paragraph{Weak Labels} GMA aims at classifying normal/ abnormal movements for complete trials/ sessions, and it normally relies on intensive observations of rigid and characterised body motions to match a predefined score sheet (cf.\ Sec. \ref{sec:background:gma}).
Existing automated approaches \cite{2012_Ubicomp_Infant}  rely on expensive expert annotations, such as the Cramped-Synchronised General Movements (CSGM) \cite{Ferrari:2002gb}. 
Such a paradigm requires sample-wise annotations and is therefore less feasible for the purpose of building up a large-scale training set. 
In fact, the number of qualified annotators is also very limited. 
One must undergo special training before being able to provide reliable annotation. 
The process is  time consuming, and the annotator needs to mark the start and end time on the sequences when CSGM occurs, which often suffers from severe inter-subject variability. 
In our case, we only have data with realistic, hence sparse trial-wise annotation (TD vs AM).
We need to train a system that not only can classify an unseen trial as either TD or AM, but also be able to indicate the moments when, for example, patterns of abnormality occurred (i.e., to identify wAM).   

\noindent
\paragraph{Generalisation}
Even though we underwent an extensive and time-consuming data collection process (including participant recruitment, logistics for the session, actual data recording and cleaning, and data annotation) the resulting data set is, strictly speaking in machine learning terms, still small-scale, which poses substantial pressure and constraints on the analysis methods.
Therefore, the second challenge is to prevent the model from overfitting by improving its generalisation capabilities for small-scale training datasets.

%% file: algorithm.tex

\subsection{Automated GMA through Analysing Accelerometer Data}
\label{sec:algorithm}

\label{sec:algorithm:overview}
\begin{figure}[t]
	\centering
	\includegraphics[width=0.98\textwidth]{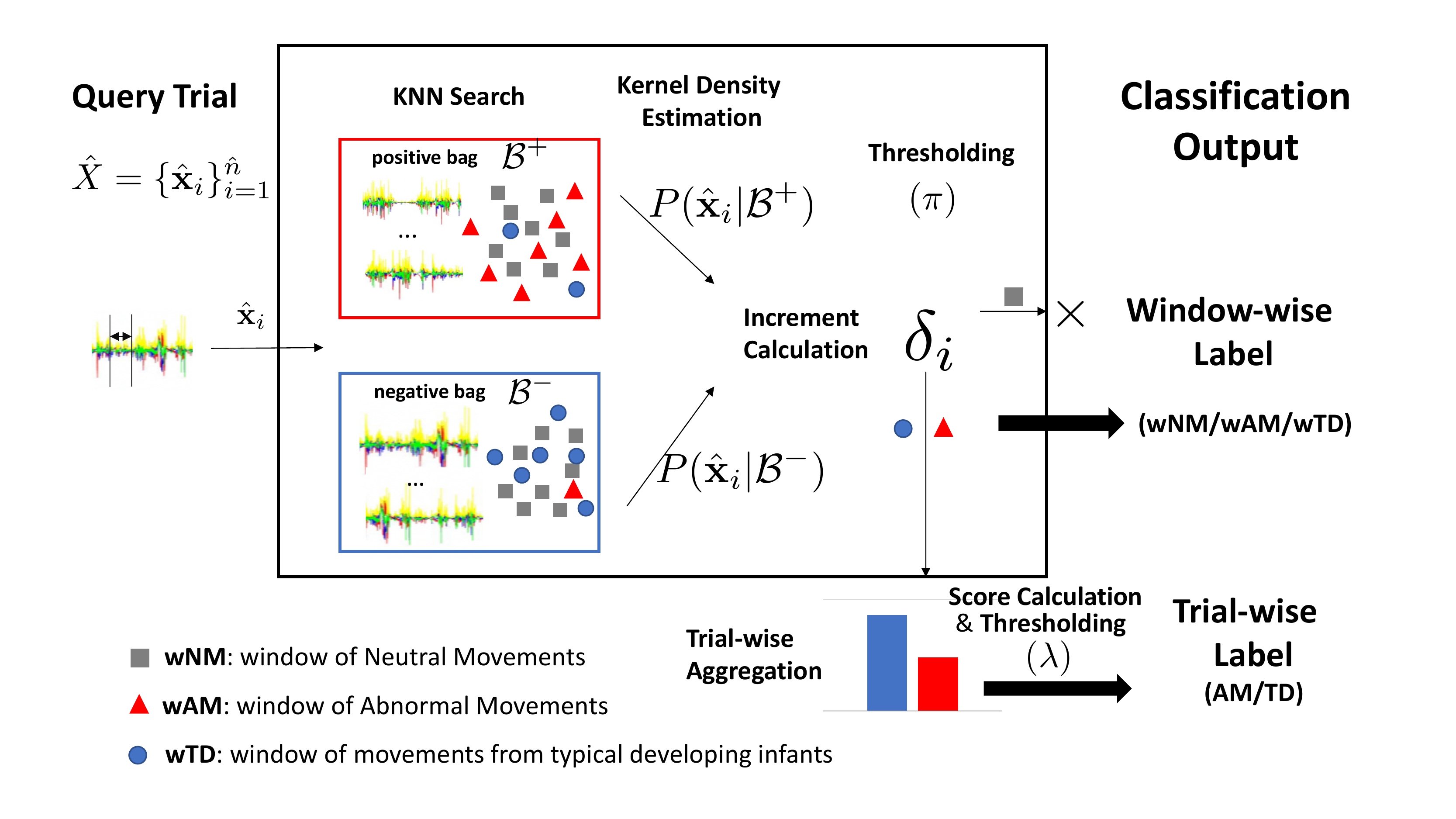}
	\caption{\label{fig:workflow}Overview of the proposed system for automated general movement assessment. See text for explanation.}
\end{figure}

\noindent\paragraph{Problem Formalisation}  
As mentioned in Sec.\ \ref{sec:methodology:challenges}, although we have trial-wise annotation, we are facing a weak label problem in modelling the system.
That is, there is no label information for each window/ frame within the trials, which may serve as important clues for final classification decision. 
In this work, we define (unseen) window labels as: 
 \begin{enumerate}
	\item wAM: window of abnormal movement, as defined in Table. \ref{tab:GMA}, which are unlikely performed by TD infants.
	\item wTD:  window of movements of a typical developing infant that cannot easily be performed by AM infants.
	\item wNM: window of neutral movement, which are the common movements that can be easily performed by both AM/TD infants.
\end{enumerate}
Based on this, within any query trial, we can assess each window independently, and aggregate all corresponding scores for the final trial-level decision. 
Fig.\ \ref{fig:workflow} illustrates the general framework of our system, visualising the two-stage assessment process, window-wise classification (wAM/ wTD/ wNM), followed by overall trial-wise classification (AM/TD).
In what follows we will provide the details of the developed analysis approach.

Raw signals of each window can be represented using low-dimensional PCA features $\mathbf{x}\in \mathbb{R}^d$ ($d=100$ in this work), and in this case the trial $X_i$ can be expressed as a collection of independent windows/ instances such that $X_i=\{\mathbf{x}_1, ..., \mathbf{x}_{n_i}\}$ with a trial-wise label $y_i\in\{0, 1\}$ indicating AM (1) or TD (0). 
In the inference stage, for query trial $\hat{X}$, we aim to predict its trial-wise label $\hat{y}\in\{1,0\}$ by accumulating all the window-wise predictions:

\begin{equation}
\centering
\hat{y}=\arg\max_{y\in\{0, 1\}} \prod^{\hat{n}}_{i=1}P(y|\hat{\textbf{x}}_i).
\end{equation}

However, owing to the lack of window-wise annotation, it is not appropriate to apply existing machine learning algorithms directly to this task.
To address this issue, in this work we define three intermediate patterns (wAM, wTD, wNM), which indicate discriminative information (wAM, wTD) to be aggregated, or redundant information (wNM) to be filtered out.
Since the discriminant movement patterns may only occur sporadically for the whole trial, it is  essential to reduce the redundant neutral movement information.
To achieve this, we propose a framework named Discriminative Pattern Discovery (DPD), which can automatically classify each window/ instance within the trial, yielding improved trial-wise classification.

\noindent\paragraph{Discriminative Pattern Discovery} 
In the field of machine learning, there are several ways of dealing with the aforementioned weak labelling problem. 
One of the most popular ways is to treat it as a Multiple Instance (MI) learning problem \cite{1997_MI}. 
In MI, instead of receiving a label for each data instance, labels are given through bags, \textit{i.e.} trials in our case. 
Specifically, MI assumes a bag is positive if it contains at least one positive instance, and is negative otherwise.
Such an assumption is similar to our problem since trials of AM infants can be modelled as a positive bag including at least one positive instance (e.g., wAM), and many negative instances.
However, such presence-based MI approaches cannot be applied directly to our problem due to some key violations to its assumption. 
For example, MI assumes that a negative bag cannot have positive instances, whereas, in our case, even TD infants (negative trials) may show occasional abnormal behaviour at times. 
As an alternative, the count-based Generalised Multiple Instance (GMI) learning was previously proposed \cite{2004_GMIL}. 
It shows a certain level of tolerance to the presence of positive (negative) instances in negative (positive) bags. 
It requires a maximum as well as a minimum number of instances of a certain concept in a bag. 
For example, a positive bag may include at least a pre-defined number of positive instances, while at most a pre-defined number of negative instances  \cite{2004_GMIL}.
We adopt and extend the idea of GMI. 
The key idea of our DPD procedure is to compute the \textit{increment} of each instance and classify it into three classes rather than two.
Specifically, for each instance instead of classifying it as positive or negative before counting the frequencies for a bag-level decision,   
 we identify its movement patterns into three pre-defined classes (or three distributions) --- wAM (or $\mathcal{A}$),  wTD (or $\mathcal{R}$), wNM (or $\mathcal{O}$)
 --- followed by an aggregation on the corresponding increments/ scores.
Our assumption is that all the instances can be drawn from these three distributions $\mathcal{A}, \mathcal{R}$, and $\mathcal{O}$. 
The classification based on DPD depends on the "soft" score proportion rather than a "hard" presence/ absence criteria in conventional MI and GMI approaches. 
For example, TD infants can have brief periods of apparently abnormal movement. 
Only if the frequency and occurrence (in terms of increment/ score) of these observations exceed the thresholds, the trial is then considered as AM. 
Accordingly, we design two \textit{Discriminative Pattern Discovery} (DPD) rules: 

\begin{enumerate}
	\item Rule $\pi$: Detecting instances drawn from the distribution of $\mathcal{O}$ (i.e., wNM ) and suppress their consequential weights.
	\item Rule $\lambda$: Learn a proportion of instances drawn from $\mathcal{A}$ (i.e., wAM), $\mathcal{R}$ (i.e., wTD) to differentiate AM/ TD trials.
\end{enumerate}

Formally, we put all instances of positive (AM) and negative (TD) trials into two bags \{$\mathcal{B}^+$, $\mathcal{B}^-$\}. 
For each instance in a query trial $\hat{\bm{x}}_i\in\hat{X}$, we are interested classifying it as wAM (i.e., from $\mathcal{A}$), wTD (i.e., from $\mathcal{R}$), or wNM (i.e., from $\mathcal{O}$).
This is measured by the increment function $\delta_i$ defined using log-odds:

\begin{equation}
\delta_i=\log\frac{P(\mathcal{B}^+|\hat{\mathbf{x}}_i)} {P(\mathcal{B}^-|\hat{\mathbf{x}}_i)},
\label{delta1}
\end{equation}

\noindent which can measure to what extent an instance can differentiate the AM/TD trials.
Given $\delta_i$, we can perform instance-level classification via rule $\pi$:

\begin{eqnarray}
\hat{\mathbf{x}}_i\in\left\{
\begin{aligned}
&\mathcal{A},&~&\text{if}~\delta_i>\pi\\
&\mathcal{R},&~&\text{if}~\delta_i<-\pi\\
&\mathcal{O},&~&\text{otherwise}.
\end{aligned}
\right.	
\label{pi}
\end{eqnarray}

\noindent Note that instances drawn from $\mathcal{A}$ or $\mathcal{R}$ provide discriminative information, while instances drawn from $\mathcal{O}$ reflect neutral movements which are redundant. 
Intuitively, if discriminative instances are outnumbered by neutral ones, that may lead to unreliable predictions.
By applying rule $\pi$, 
we can classify $\hat{\mathbf{x}}_i$ into the aforementioned 3 classes, and apply  
a simple masking operation such that $\hat{h}_i=0$ if $\hat{\mathbf{x}}_i\in\mathcal{O}$, and $\hat{h}_i=1$ otherwise. 
Based on this masking operation, we can easily filter out the non-discriminative movement patterns (wNM) for the query trial, leaving discriminative ones for the trial-wise assessment. 
That is, the overall trial $\hat{X}$ can be assessed by aggregating all discriminative patterns, and rule $\lambda$ can be applied for the final trial-wise classification (TD/ AM). 

\begin{equation}
	\hat{y}=\left\{
	\begin{aligned}
		1, &\ &~&\text{if} \  \frac{1}{\sum_{i=1}^{\hat{n}}\hat{h}_i}\sum_{i = 1}^{\hat{n}}\hat{h}_i\delta_i> \lambda \\
		0, &\ &~&\text{otherwise}.
	\end{aligned}
	\right.
	\label{lambda}
\end{equation}

\noindent\paragraph{Kernel Generalisation}
For machine learning based systems, generalisation capability to unseen environments is of great importance.
In our case, considering the fragility of the population and the constraints imposed on assessment procedures on infants, collecting large amounts of training data is expensive. 
So far, our dataset contains 161 validated trials that vary with regard to age, gender, terms, etc. 
These factors result in large variability in movement patterns and thus may degrade the model performance for unseen environments. 
In this case, sophisticated algorithms may suffer from overfitting problems with poor generalisation.
Our solution for this problem follows the GMI paradigm \cite{1997_MI,2004_GMIL}, which assumes that each instance is drawn from smoothed distributions by learning effective kernel functions. 
A typical way is to compress the training set into a very small codebook by unsupervised clustering. 
Each cluster can be viewed as a representative pattern that summarises surrounding samples into a compact kernel space so that the large variability can be reduced.
However, such approaches often suffer from the \textit{quantisation problem}, i.e., the inherent inaccuracy of representation when mapping the original data to a few representatives.
For example, we have roughly 70,000 instances which shall be clustered into, e.g., 200-1,000 representative patterns, which inevitably leads to loss in precision of the representation. 
As will be shown in the experiments, although quantifying specific instances to clusters may mitigate variability, we may also sacrifice discriminative information substantially. 
In addition, existing clustering algorithms (e.g., K-means), often suffer from the randomness and low efficiency. 
These problems motivate us to design a better generalisation algorithm so as to circumvent the quantisation issue. 
Now we aim to model the increment $\delta_i$ for query instances, and via Bayesian theorem Eq. (\ref{delta1}) can be  reformulated as:

\begin{eqnarray}
\delta_i&=& \log\frac{P(\hat{\mathbf{x}}_i|\mathcal{B}^+)P(\mathcal{B}^+)}{P(\hat{\mathbf{x}}_i|\mathcal{B}^-)P(\mathcal{B}^-)} \nonumber  \\
&=&\log P(\hat{\mathbf{x}}_i|\mathcal{B}^+)-\log P(\hat{\mathbf{x}}_i|\mathcal{B}^-)+ c, 
\label{delta_2}
\end{eqnarray}

\noindent
where $c=\log\frac{P(\mathcal{B}^+)}{P(\mathcal{B}^-)}$ is a prior constant that can be estimated by the proportion between numbers of AM and TD trials. 
With this we now aim to estimate the density $P(\hat{\mathbf{x}}_i|\mathcal{B})$ for both bags.
We employ an online algorithm based on non-parametric probability density estimation \cite{2001_PDF}. 
The primary idea is to only consider task-relevant instances at the test time instead of blind quantisation at the training stage. 
Given a query instance, the algorithm first searches for a number of similar instances (e.g., via k-nearest neighbour method) in a certain training bag, which can be used to smooth the noise and variability accordingly. 
This process is supported by the theory of long-tail characteristic of high-dimensional features, i.e., instances that are far away from the query data in the feature space often make less contribution to the density estimation. 
Specifically, we can estimate the kernel density of the query data by taking the integral of these surrounding relevant instances, such that:

\begin{equation} \label{density}
P(\hat{\mathbf{x}}_i|\mathcal{B})=\frac{1}{k}\sum_{j=1}^{k} \mathcal{K}(\hat{\mathbf{x}}_i-\bm{NN}_j)
\end{equation}

\noindent
where $K: \mathcal{X}\times \mathcal{X}\rightarrow\mathbb{R}$ is a Parzen kernel function that is non-negative and integrates to 1. 
Without loss of generality, we use a Gaussian kernel: $\mathcal{K}(\hat{\mathbf{x}}_i-\bm{NN}_j)=\exp(-\|\hat{\mathbf{x}}_i-\bm{NN}_j\|^2)$. 
Note that $\bm{NN}_j$ is the $j^{th}$ of top $k$ nearest neighbour search results of $\mathbf{x}_i$ for bag $\mathcal{B}$. 
With Eqs.\ (\ref{delta_2}) and (\ref{density}) the increment function $\delta_i$ can be written as: 

\begin{equation} \label{delta_3}
\delta_i=\log\frac{1}{k}\sum_{j=1}^{k} \exp(-\|\hat{\mathbf{x}}_i-\bm{NN}^+_j\|^2)-\log\frac{1}{k}\sum_{j=1}^{k} \exp(-\|\hat{\mathbf{x}}_i-\bm{NN}^-_j\|^2) + c.
\end{equation}

\noindent
As mentioned before, $c$ is a constant and can be merged into the $\lambda$ rule (for the trial-wise thresholding) and thus does not need explicit computation. The overall algorithm is summarised in Algorithm \ref{alg}.


\begin{algorithm}[t]
 \KwIn{Positive and Negative Bags $\mathcal{B}^+,\mathcal{B}^-$; Query trial $\hat{X}=\{\hat{\textbf{x}}_1,\cdots,\hat{\textbf{x}}_{\hat{n}}\}$; Empirical Hyperparameters $k$, $\pi$, and $\lambda$;}
 \KwOut{Prediction $\hat{y}$.}
 $h=0$\;
 \For{all query instances $\hat{\mathbf{x}}_i$ $\in$ $\hat{X}$}{
		Compute $k$ Nearest Neighbours of $\hat{\mathbf{x}}_i$ in $\mathcal{B}^+$;\\
		Compute $k$ Nearest Neighbours of $\hat{\mathbf{x}}_i$ in $\mathcal{B}^-$;\\
		Compute the increment $\delta_i$ using Eq.(\ref{delta_3});\\

                \eIf{$|\delta_i|$ < $\pi$}{
                     $\hat{h}_i$ = 0\;
                      }{
                        $\hat{h}_i$ = 1\;
                        $h \leftarrow h + 1$\;
                         }
                  $\Delta \leftarrow \Delta + \hat{h}_i \delta_i$\;
      }
 \eIf{$\frac{1}{h} \Delta > \lambda$}{
                                                               $\hat{y}$=1\;
                                                              }{
                                                               $\hat{y}$=0\;
                                                              }
\caption{Discriminative Pattern Discovery with Kernel Generalisation}
\label{alg}
\end{algorithm}

%% file: experiments.tex
\section{Experimental Evaluation}
\label{sec:experiments}
Our evaluation design is inspired by the GM Trust Course on the Prechtl Assessment of General Movements \cite{2002_CSGM}. 
We first compare our approach to a list of baseline methods as the ablation experiment. 
The evaluation is based on our collected dataset as introduced in Sec.\ \ref{sec:methodology:data_acquisition}.
We also provide an in-depth analysis of the effects the algorithmic details, such as the implementation and hyper-parameters, have on the overall performance.

\subsection{Settings}
\label{sec:experiments:settings}
Our experiments are based on 10-fold random split (149/12) cross-validation. 
That is, every fold contains 12 trials as test, which is in line with the Prechtl Assessment test. 
Over the remaining training trials, we further use 10-fold (137/12) cross validation to find the optimal hyper-parameters. 
This setting focuses on theoretical model comparison and thus considers the participant as one type of covariate factor together with age, head position, outliers, etc. 

\subsection{Baselines}
\label{sec:experiments:baselines}
We compare the proposed DPD method to a range of alternative, conventional machine learning methods.
Despite popularity, existing deep learning techniques (e.g., \cite{2015_Yang_1dCNN_HAR,2016_roggen_ConvLSTM,2016_Hammerla_IJCAI,2017_Guan_LSTMensemble,Zeng:2014tk}) are not applicable to this problem because these methods need training data with sample/window-wise annotations, which are not available in our case. 
The baselines are as follows: 
\begin{description}
	\item [KNN:] 
	Since our kernel embedding is based on $k$-Nearest Neighbour (KNN) search, an intuitive baseline is KNN itself. 
	Specifically, we first search $k$ nearest neighbours of every query instance $\hat{\mathbf{x}}_i$ in the training set (i.e., AM/TD trials). 
	The instance is classified as AM or TD depending on the $k$ nearest instances' trial-wise labels. 
	The trial-wise classification is then performed via majority voting on classified instances.
	
	\item [SVM:]
	Here all the instances in the AM trial are presumed to share the same label--AM--despite a fairly large portion of them may not present significant abnormal movements. 
	Considering SVM is based on learning the support vectors between two distinctive distributions, we concern whether such a mechanism can be readily applied. 
	We implement the SVM using LibSVM toolbox with a Gaussian kernel. 
	
	\item [GMI-GEN:] 
	One of our key arguments for \textit{Kernel Generalisation} is that clustering-based approaches may degrade the performance due to quantisation and loss of original discriminative information. 
	In order to validate this hypothesis we run Generalised Multiple Instance-Generative (GMI-GEN) baseline experiments. 
	The GMI-GEN model is implemented by initially quantising the training set into a compact set of representative clusters using K-means, while we keep 
	other details exactly the same as our proposed method. 
	
	\item [Ours (no DPD):] 
	The key difference of our method to conventional MI approaches is that we classify instances into 3 classes, and suppress the redundant wNM. 
	Such a procedure refers to the hyper-parameter $\pi$ in Alg.\ \ref{alg}. 
	In this baseline, we take all of the instances (also including the redundant wNM) into account without applying the $\pi$ rule. All other parameters are kept exactly the same as for our proposed method.
	
	\item [Prechtl's Standard:] 
	To achieve a basic certificate, a human observer needs to correctly classify at least 75\% (in our case 9/12) of randomly selected AM/ TD trials. 
	Our evaluation simulates such a ``random-12'' setup as it is used for clinician training and assessment, and thus the results are directly interpretable for practical scenarios.
\end{description}

\subsection{Evaluation Metrics}
\label{sec:experiments:metrics}
We use accuracy as the main metric to compare different baselines to our approaches. 
Due to the (slightly) imbalanced distribution of AM and TD trials (see Table \ref{tab_participants}), we also adopt other popular metrics based on True Positive (TP), False Positive (FP), True Negative (TN) and False Negative (FN) classifications.
All evaluation metics used are summarised in what follows:
\begin{description}
	\item [Accuracy:] Counts the number of correctly classified trials, normalized over the total number of trials. 
	
	\item [Sensitivity (Recall):] TP/(TP+FN), measures the proportion of positives that are correctly identified as AM.
	
	\item [False Positive Rate:] FP/(FP+TN) quantifies to what extent TD infants are misclassified as AM.
	
	\item [Specificity:] TN/(FP+TN), also known as true negative rate, measures the proportion of negatives that are correctly identified as TD.
	
	\item [Precision:] TP/(TP+FP) is the fraction of AM trials among the overall positive predictions.
\end{description}

\noindent
We also provide confusion matrices and Receiver Operating Characteristic (ROC) curves for further discussion.

\subsection{Main Results}
\label{sec:experiments:main_results}
Table \ref{tab:comparison} shows the results for different baselines and our method.
It can be seen that our method consistently outperforms other methods with the highest overall accuracy (80\%). 
It is worth noting that the proposed automated assessment method is able to correctly classify AM with at least the accuracy that is required by trained human annotators  (75\%) to pass the GM exams. 
Our model can therefore provide an objective, automatically generated reference for clinical purposes. 
Next, we analyse the contributions of our approach through comparing it to baselines. 

\begin{table}[t]
	\caption{Classification accuracies for 10-fold cross-validation experiments (means and standard deviations ($\pm$))}.
	\label{tab:comparison}
	\begin{tabular}{lccccc}
		\toprule
		Method&Accuracy&Sensitivity(Recall)&False Positive Rate&Specificity&Precision\\
		\midrule
		KNN & 0.22 (0.12)&1.00 (0.00)&1.00 (0.00)&0.00 (0.00)&0.22 (0.12)\\
		SVM & 0.79 (0.11)&0.00 (0.00)&0.00 (0.00)&1.00 (0.00)& -\\
		GMI-GEN & 0.32 (0.17)&0.73 (0.33)&0.80 (0.23)&0.20 (0.23)&0.20 (0.12)\\
		Ours(No-DPD) & 0.70 (0.10)&\textbf{0.88 (0.16)}&0.32 (0.13)&0.68 (0.13)&0.43 (0.20) \\ \hline
		Ours & \textbf{0.80 (0.13)}&0.70 (0.35)&\textbf{0.13 (0.11)}&\textbf{0.87} (0.11)&\textbf{0.57 (0.27)}   \\
		\bottomrule
	\end{tabular}
\end{table}

The first two baselines have lowest recognition performance. 
KNN predicts all of the query trials as AM, whereas SVM goes to the other extreme with all TD predictions. 
We ascribe the failure of KNN to poor generalisation because test instances of new trials may be very dissimilar to the training trials. 
Similarly, the performance of the SVMs shows its low generalisation -- despite the fact that we employ a Gaussian kernel that explicitly aims to capture local distributions for better generalisation.
Another reason for the SVM's failure may be the weak labelling problem, and in the experiments we presume all the instance labels share the same one with the trial.
The experimental results suggest that traditional machine learning methods cannot be used directly in our scenario, since they cannot cope well with the weak labelling problem.

GMI-GEN is a variant of our method, and it initially performs clustering for compact representation. 
We notice that its recall is reasonable, yet the false positive rate is very high (0.8). 
Generally, GMI-GEN is significantly worse than our approach, since the discriminative information may be lost during the clustering process. 
The second best approach (i.e., No-DPD) is the variant of our algorithm that does not eliminate contribution from the instances drawn from $\mathcal{O}$ (i.e., wNM). 
The accuracy is only 10\% lower than our approach, and the recall is even higher. 
We observe that its main weakness is the specificity and False Positive Rate (19\% lower and 19\% higher, respectively). 
These results suggest the importance of suppressing redundant wNM via DPD's $\pi$ rule.

\begin{figure}
	\centering
	\includegraphics[width=1\textwidth]{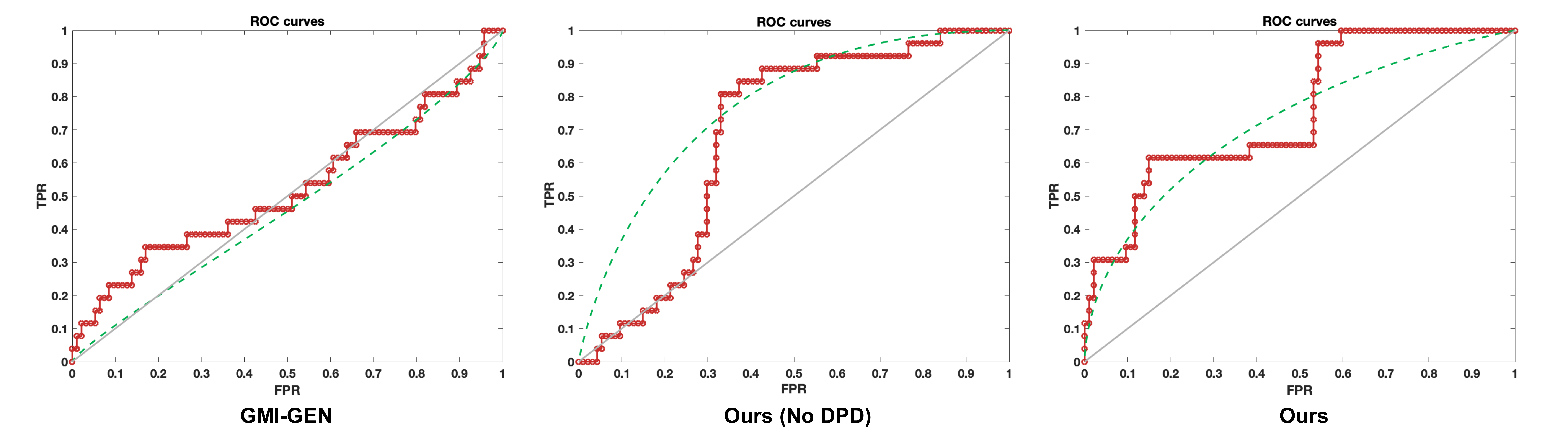}
	\caption{\textbf{\label{fig:roc}} ROC curves of key baselines.}
\end{figure}

\begin{figure}
	\centering
	\includegraphics[width=1\textwidth]{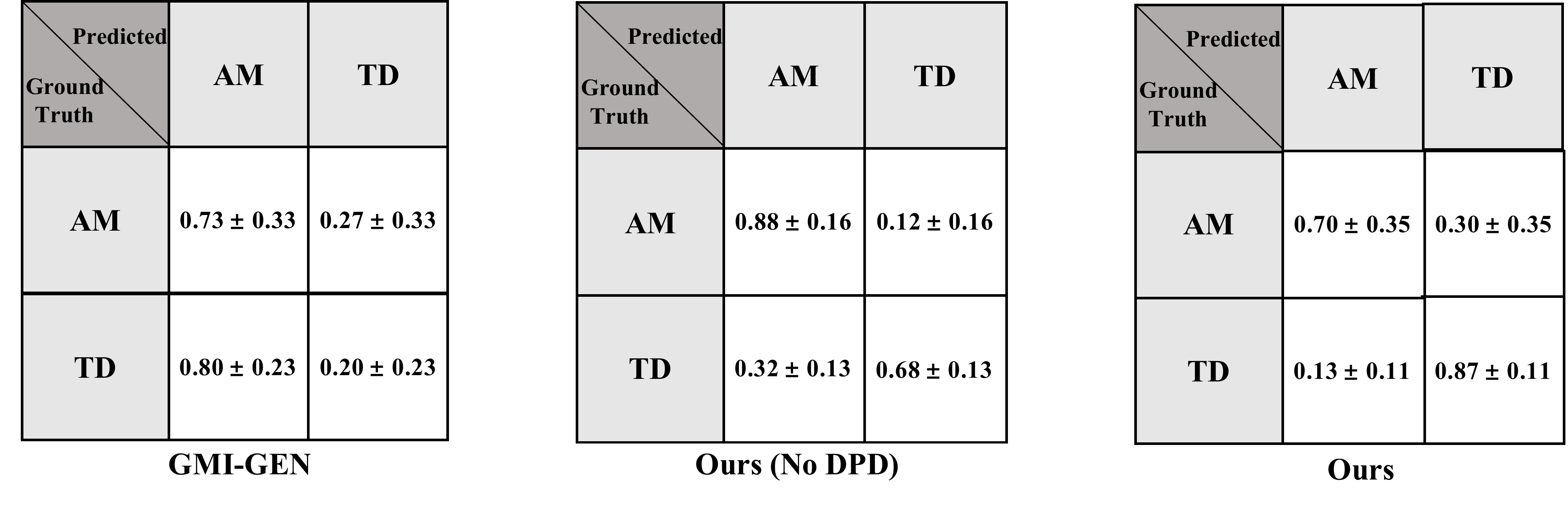}
	\vspace*{-1.5em}
	\caption{\textbf{\label{fig:cm}} Confusion Matrices of key baselines (means and standard deviations ($\pm$)).}
	\vspace*{-1em}
\end{figure}

Fig. \ref{fig:roc} shows the ROC curves of the best three approaches (dashed curves denote smoothed trend lines). 
GMI-GEN performs closely to random guess but gets better when the recall is low (higher TPR/ FPR). 
The overall Area Under Curve (AUC) of No-DPD is slightly higher than that of DPD. 
However, the DPD approach converges to perfect TPR earlier (roughly 0.6 compared to 0.85 of No-DPD), which means that this approach is more stable and can provide more reliable predictions with lower costs.  
Confusion matrices in Fig.\ \ref{fig:cm} for GMI-GEN, No-DPD, and DPD provide class-based insights that allow to assess the trade-off between AM and TD predictions.

\subsection{Technical Analysis}
\label{sec:experiments:detailed_analysis}
Classification experiments as discussed in the previous section demonstrate the effectiveness of our DPD framework for successfully classifying trials into AM/ TD.
These results are encouraging for our overall research agenda and vision towards a reliable automated system that enables population-wide screening of infants (on AM detection) for early diagnosis.
In addition to these clinically oriented evaluations it is also of importance to further analyse the technical details of our method, which provides insights into the wider capabilities of the approach -- potentially beyond the dataset analysed in this paper. 
Specifically, we are interested in how the hyper-parameters may affect the overall performance of our method.
We will also provide some qualitative analysis on the recognition results.


\subsubsection{Hyper-parameters} 
\label{sec:experiments:detailed_analysis:hyperparameters}
Our method has three main hyper-parameters.
In what follows we will explain the role these hyper-parameters play and present evaluation results that provide the basis for quantitative comparisons.

\begin{figure}
	\centering
	\includegraphics[width=0.7\textwidth]{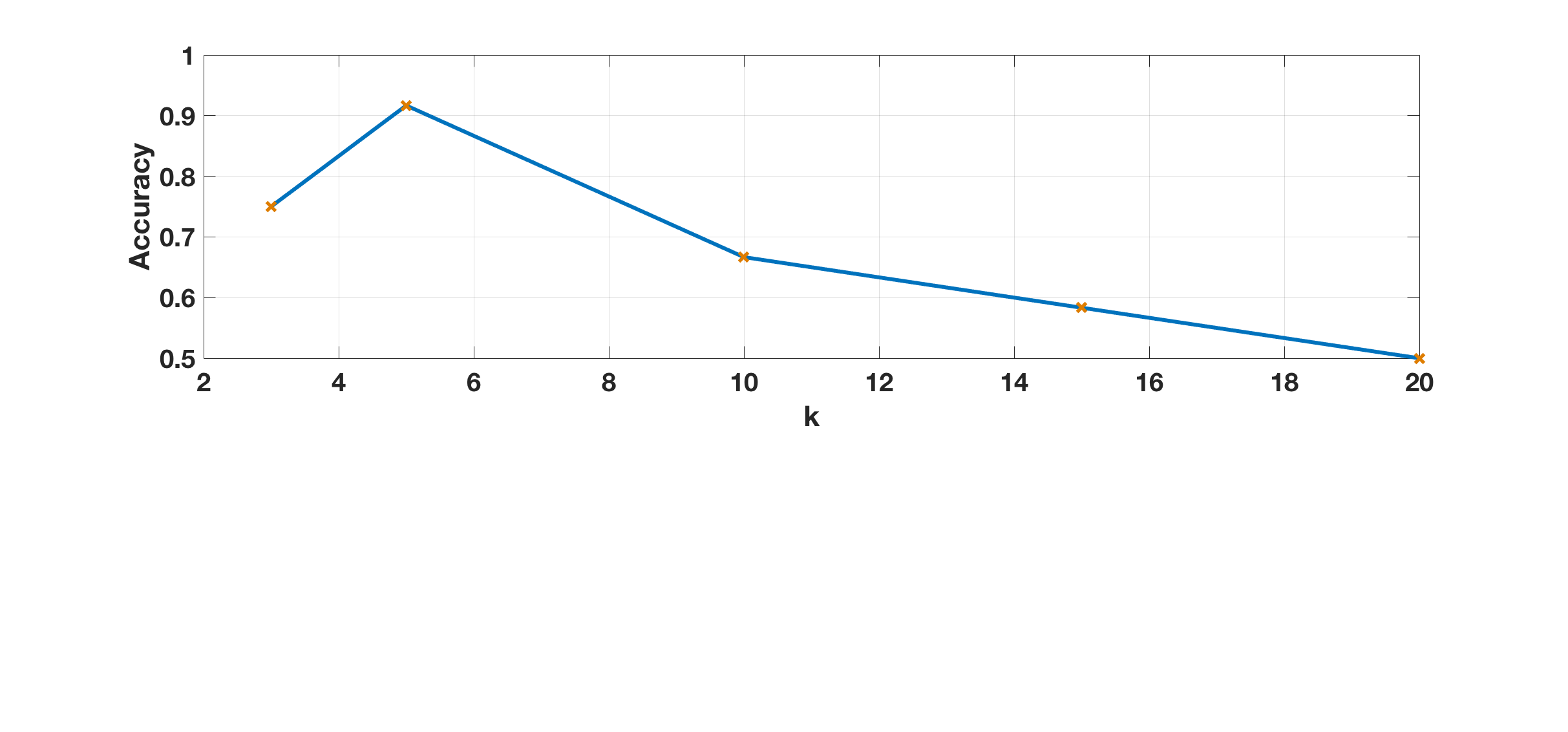}

	\vspace{-15ex}
	\caption{\textbf{\label{fig:k}} Performance with respect to the number $k$ used in k-NN.}
	\vspace*{-1em}
\end{figure}

\begin{figure}
	\centering
	\includegraphics[width=0.7\textwidth]{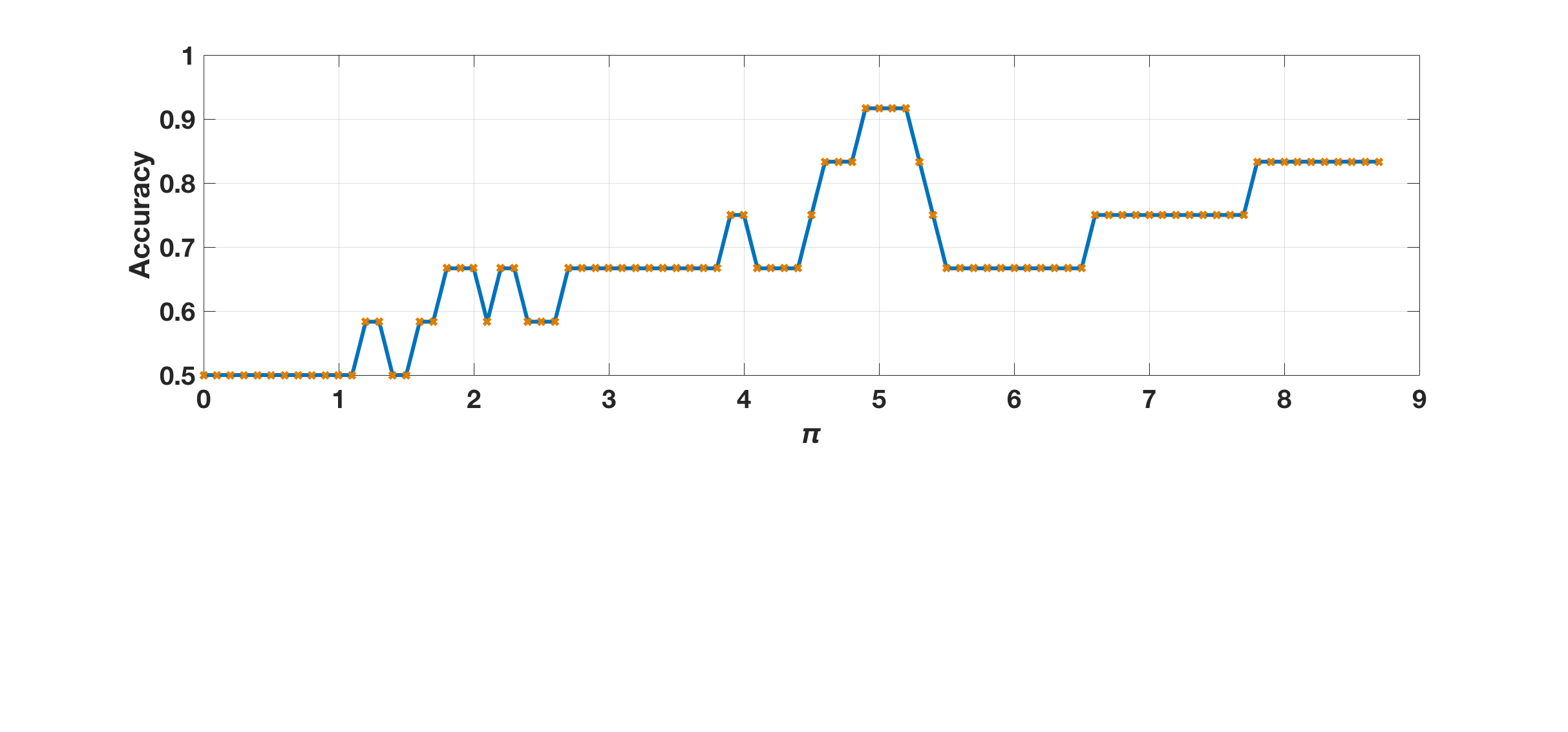}
	\vspace{-15ex}
	\caption{\textbf{\label{fig:pi}} Performance with respect to $\pi$.}
	\vspace*{-1em}
\end{figure}

The number $k$ used for $k$-NN based analysis generally dictates the smoothness of the decision boundary and thus has a direct impact on classification performance (as shown in Fig.\ \ref{fig:k}). 
We can see a smaller value of $k$ yields better performance. 
$\pi$ is used as a threshold to identify discriminative patterns (wTD and wAM) and redundant instances (wNM), as shown in Eq. (\ref{pi}).  
In Fig.\ \ref{fig:pi}, we can see better results can be achieved by using a relatively larger value of $\pi$, which can filter out most of the non-discriminative wNM instances, since the corresponding increments are within the range of $-\pi<\delta_i<\pi$. 
However, performance drops when $\pi$ goes larger, since in this case more discriminant instances (i.e., wAM and wTD) will be classified as redundant wNM to be suppressed. 
The identified wAM/wTD instances may provide important clues (as shown in Fig. \ref{fig:window}) for final trial-wise classification. 
The hyper-parameter $\lambda$, on the other hand, is used to threshold the aggregated scores (from wTD and wAM instances of a trial) in order to identify the whole trial as AM/ TD, as shown in Eq. (\ref{lambda}). 
We can see that $\lambda$ is not very sensitive within the range of $[-8, 0]$, and it indicates that wAM instances (with positive increments $\delta_i >\pi$) are rare events (compared with wTD with negative increments $\delta_i <-\pi$) for AM trials.
For these three key hyper-parameters, in our experiments we set $k=5$ and adaptively choose the optimal $\lambda$ and $\pi$ via cross-validation corresponding to each test split.

\begin{figure}
	\centering
	\includegraphics[width=0.7\textwidth]{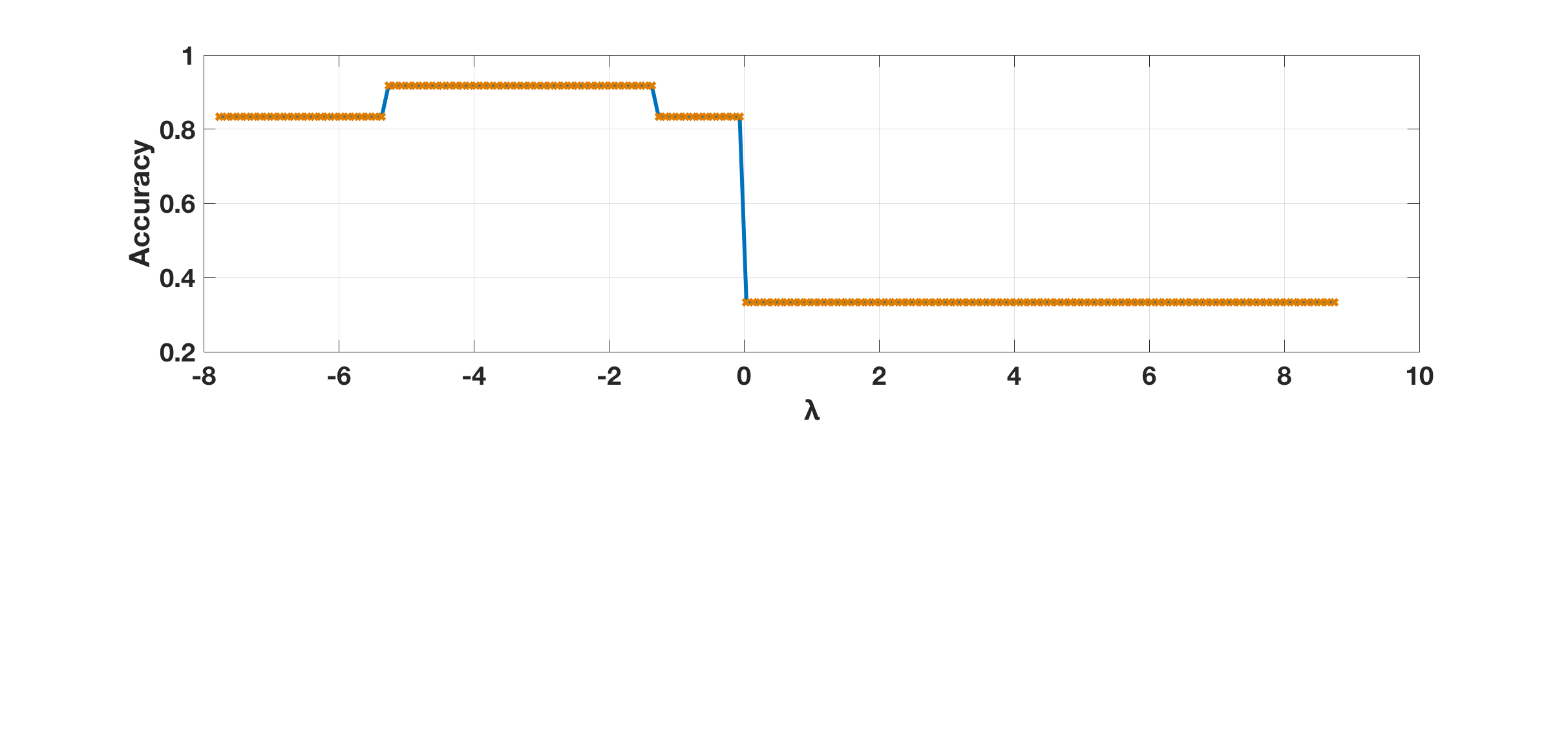}
	\vspace{-15ex}
	\caption{\textbf{\label{fig:lambda}} Performance with respect to $\lambda$}
	\vspace*{-1em}
\end{figure}


\subsubsection{Qualitative Analysis} 
\label{sec:experiments:detailed_analysis:qualitative}
\begin{figure}
	\centering
	\includegraphics[width=0.9\textwidth]{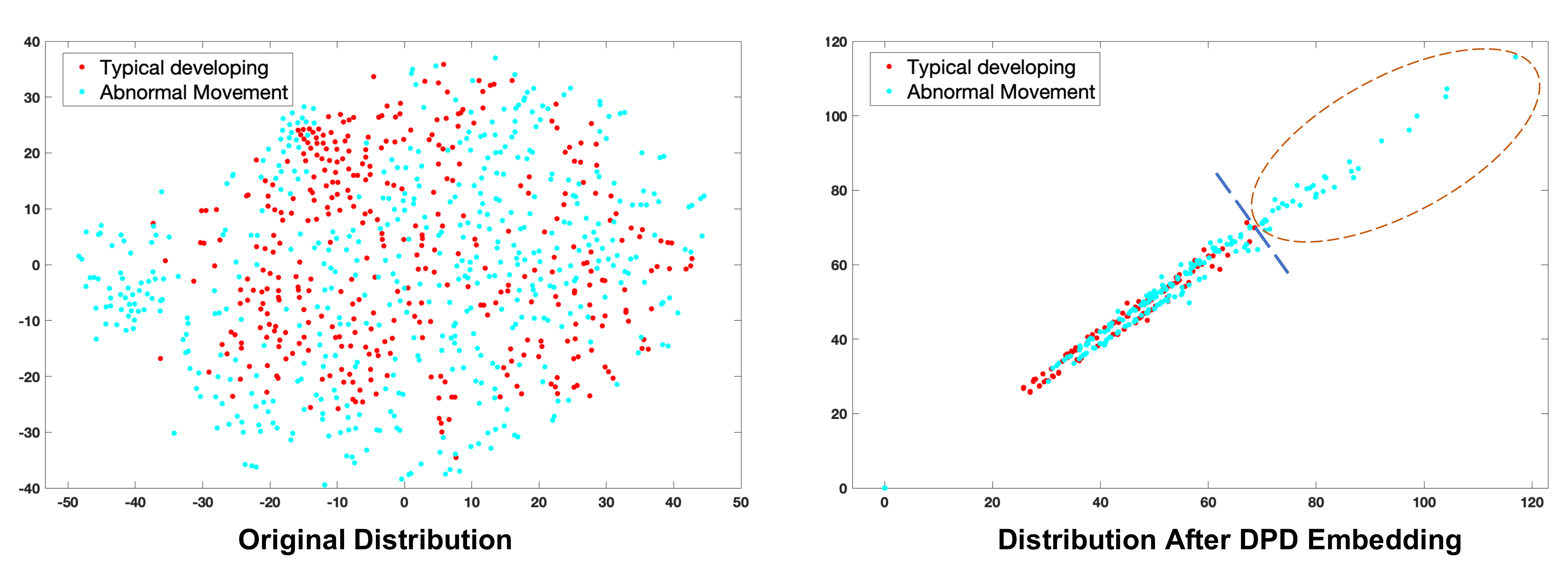}
	\caption{\textbf{\label{fig:tSNE} Qualitative study:} DPD embedding (right) leads to improved class separability and thus better classification performance (tSNE visualisation; red circle in right plot visualises kernel embedding; best viewed in colour).}
\end{figure}

In Fig.\ \ref{fig:tSNE}, we visualise the distributions of an example pair of AM and TD trials using the t-SNE method \cite{vanderMaaten2008-VDU}. 
The method projects the 100-D PCA feature point space into 2-D space enabling visual comparisons and analysis of the relations between AM and TD points. 
It can be seen that our DPD embedding leads to data distributions that are much easier to separate, resulting in improved classification performance.

\begin{figure}
	\centering
	\includegraphics[width=1\textwidth]{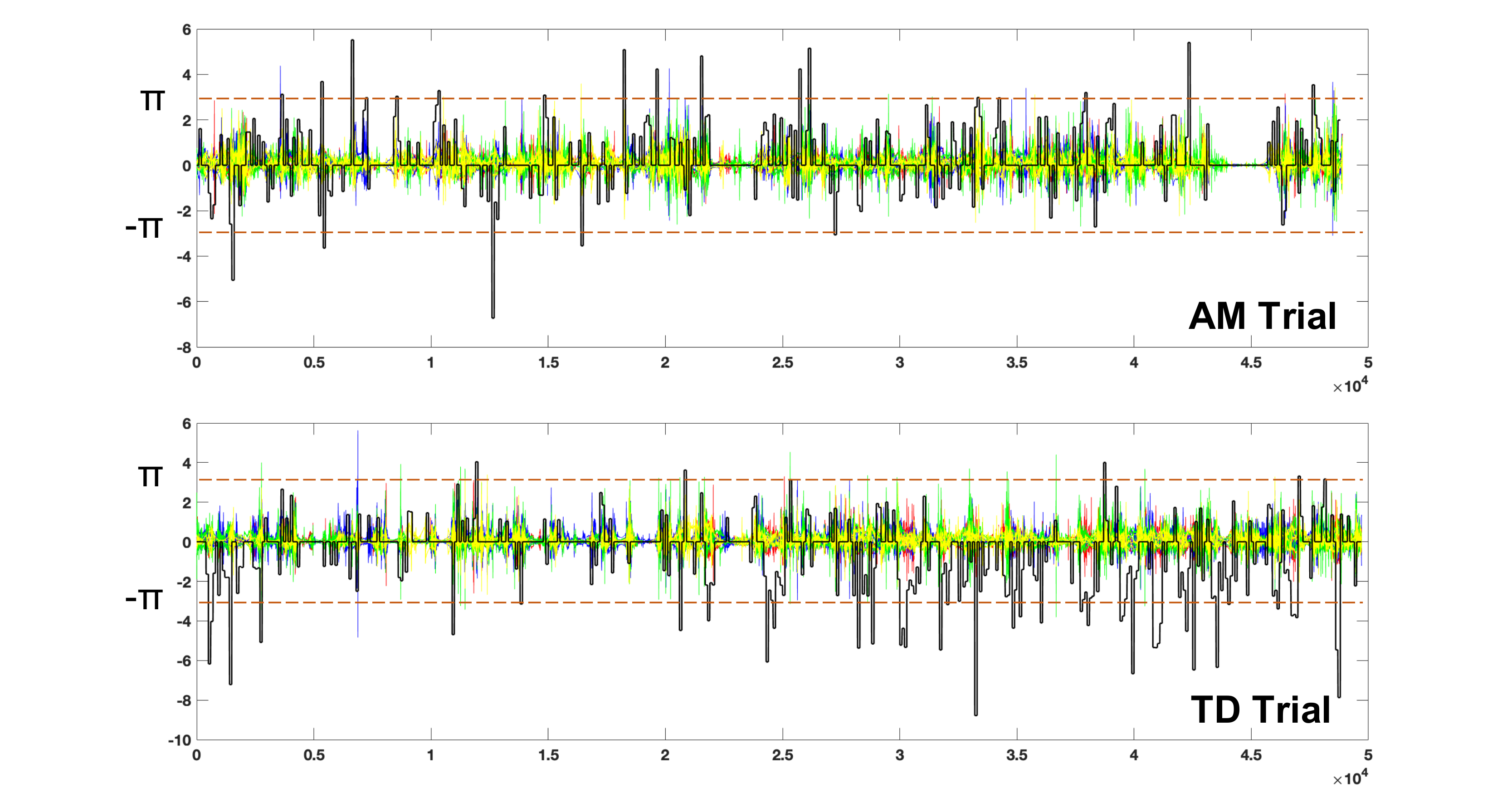}
	\caption{\textbf{\label{fig:window} Qualitative study:} Instance-wise classification that can be used to identify three pre-defined movements, i.e., wTD, wAM, and wNM. (solid black lines indicate the distribution of increment $\delta_i$ within TD/AM trials, while other colours corresponding to the raw signal).}
\end{figure}

One important advantage of our DPD framework is that we also perform instance-wise (i.e., window-wise) classification, and each instance can be measured by $\delta_i$, and classified into wAM, wTD, or wNM.  
Identifying the three types of movements is important, and the identification of wAM instances can be used as  evidences for automated screening. 
Fig.\ \ref{fig:window} visualises complete trials (TD and AM). 
Colourful sequences denote raw data while the overlaid solid, black lines show the $\delta_i$ for each instance/ window over the temporal line.
We can see for both TD/ AM sequences, the wNM instances (corresponding to $-\pi<\delta_i<\pi$) are the dominant movement patterns.
After removing the redundant wNM instances, we can clearly see the discriminant patterns wAM ($\delta_i>\pi$) and wTD ($\delta_i<-\pi$), which are used for trial-wise score calculation, and their distribution for AM/TD trials, which would provide important clues (e.g., timestamps that can be used to trace the synchronised video clips) for better behaviour understanding.

 


%% file: conclusion.tex

\section{Discussion}
\label{sec:conclusion}
\subsection{Summary}
\label{sec:conclusion:summary}
Stroke is a serious event that, if untreated, can lead to substantially adverse outcomes.
While it is relatively straightforward to diagnose a stroke in adults, a stroke before or right after birth is much harder to detect.
Such Perinatal Strokes (PS) are not uncommon but, unfortunately, many cases remain undiagnosed for too long with adverse consequences.
A large percentage of infants affected with PS will develop severe conditions such as Cerebral Palsy that have lifelong impact.
In this paper we presented an automated assessment system, which detects abnormal movements (AM) of infants with Perinatal Stroke.
Movement data are captured through miniaturised, inexpensive inertial measurement units and analysed through a novel, machine-learning based sensor data analysis method.
This data analysis approach effectively tackles a weak label problem in which only very few and rather coarse ground truth annotations are provided for model training.
In practice clinicians classify trials of, for example, ten minutes duration with regards to whether they consider the observed infant as typically developing or showing abnormal movements.
The developed Discriminative Pattern Discovery (DPD) method automatically detects relevant patterns and bootstraps effective classification models based on these.

We evaluated the effectiveness of the developed framework in a deployment study.
Data were recorded and analysed for a total of 34 infants (21 typically developing, 13 with clinically diagnosed PS -- each confirmed through MRI), each at various steps in their development after birth.
For this dataset our automated assessment system was able to correctly discriminate between TD and AM, with a higher accuracy than a GMA-trained clinician would need to achieve in order to be able to pass their course exam.
These are very encouraging results because they demonstrate that automated PS screening is possible.
Due to the low costs of the sensing hardware and the minimal effort to use these, larger scale uptake is realistic.

\vspace*{-0.5em}
\subsection{Lessons Learned}
\label{sec:conclusion:future}
The work presented in this paper is explorative, serving as the basis for our overarching endeavor of developing effective methods and tools for screening newborns for perinatal stroke.
Due to the importance of early, reliable recognition of indications of PS we ultimately aim for population-wide screening, which not only requires accuracy and reliability of the automated assessment but also robustness, straightforward deployment with minimal disruption, and wider stakeholder engagement.
Based on our experiences reported in this paper, in what follows we offer insights into lessons learned while developing our system.

\vspace*{-0.5em}
\paragraph{Sensing Infrastructure}
Without a doubt attaching any kind of equipment on newborns is and will remain a challenge. 
However, the alternative of, for example, using a camera based data capturing approach does not seem viable not only due to privacy concerns (which could be addressed) but mainly due to practical concerns related to quality of illumination, or occlusions.
Furthermore, the temporal resolution of inertial measurement units as the ones used for this paper exceeds that of standard cameras.
Direct movement sensing thus is advantageous over indirect procedures such as video-based ones.
Our sensors were remarkably robust to the extent that it was not problematic at all to reliably record movement data and to transfer them to the processing machine.
We used commercially available, inexpensive sensors that can almost be considered commodity devices that caused no substantial problems to the data collection process.
As such, this setting seems suitable for the envisioned larger scale screening efforts.

\paragraph{Data Collection}
The data collection procedure followed a relatively strict protocol, which is suitable for the clinical setting in which it took place.
Arguably, screenings at home will require a more flexible procedure as parents of newborns will experience substantial disruptions already through the arrival of their infant (often multiplied if more than one child has to be cared for). 
As such any additional activity may not be straightforward to integrate if it is too disruptive or burdensome, and thus needs to be designed and optimized very carefully.
The main concern here lies in the specialised sensor synchronisation procedure that requires some effort and would thus potentially be difficult to integrate into everyday routines outside the clinical environment.
Despite these concerns, we were pleasantly surprised that the practicalities of handling the sensing system by the clinical personnel (not the researchers) did not cause major problems. 
In fact, it was straightforward for them to attach the sensors to the limbs of the infants, which was facilitated by our bespoke wrist- and ankle-bands.
However, even with the current level of miniaturisation of wearable technology, the size of any sensing platform remains a challenge for scenarios that involve infants.
As such it is important to have the bands tightly fit around the limbs.

The infants' movements were not hindered by our data collection. 
However, this may change when the infants grow and become more mobile. 
Also, as the infants mature they will certainly become more interested in the wrist- and ankle-bands, such that the resulting playfulness may interfere with the data collection.
To mitigate such issues we may need to change the design and possibly the form-factor of the pouches.
However, the sensors itself would not need to be changed as they are small enough.
Also related to growth and maturing, the limitation to sensing on the limbs only may become more problematic.
Prechtl's GMA not only focuses on limb movements but also on movements of the head, general synchrony of movements, full posture and center of mass distributions.
Limiting data recording to limbs only would need to be extended, for example, by attaching sensors to the trunk.
Disposable accelerometers in form of patches are now available, which could be used to complement the limb-based movement recording.


\paragraph{Data Analysis}
Analysing the recorded sensor data represents a formidable challenge as elaborated on in this paper.
We have developed an effective method that recognises atypical movements of infants in clinical environments.  
The most important insights on the data analysis part of our project are related to data quality and how to address issues therein.
It turned out to be of substantial importance to de-noise the multimodal sensor data, which we have achieved through a PCA-based method.
This method not only reduced noise in the data but also eliminated redundancy in the recordings.
We preferred a data-driven approach over domain-driven manual optimisation because at this time it is not yet clear what parts of the recorded data are most informative for the analysis and thus a holistic approach such as PCA seemed more suitable (for this pilot study) than dedicated filtering or feature engineering approaches.

Infants do not move very fast and as such one would assume that a low sampling rate would be sufficient for the inertial measurement units to capture the relevant movements (according to the Nyquist theorem from general signal processing). 
However, even though the frequency of gross limb movements of the infants certainly does not exceed a few Hz and thus sampling rates between 10 and maximally 20 Hz would be sufficient for capturing, the higher sampling rates of our sensors were beneficial. 
In addition to gross limb movements it is the subtle aspects such as jitters and small jerky movements that catch a human annotator's attention and thus contribute to the overall assessment.
Our high sampling rates (100Hz) were beneficial for  the overall analysis (which is in line with previous work on health related movement analysis \cite{2012_Plotz_SB}).

\subsection{Way Forward: Population-wide PS Screening}
\label{sec:conclusion:realworld}
Our assessment setup is scalable and can thus serve as basis for population-wide PS screening.
Screening could be undertaken on the delivery suit, and as part of the routine visit by health visitors, and midwives, or even by the parents at home. 
The sensors, which are paired with smartphones, are small enough to not hinder nor harm the infant and as such movement recording is straightforward.
Opportune moments for recording and assessment would be 
anytime the infant is quiet yet alert.
These moments could either be indicated manually by the parents (for example, through a companion app on their phones) or through an automated approach.
Through this, the data collection and automated assessment can be conducted in a more frequent manner, i.e., continuous monitoring, in contrast to the monthly assessment in this work.

Given the rapid development of infants after birth, substantial changes in their general movement patterns are to be expected. 
With continuous monitoring scheme, such generic changes in movement patterns that are not related to a potential perinatal stroke could be incorporated into the modelling process.
For example, the models could be continuously adapted by using both newly recorded data and the predictions that are automatically generated by our system (and potentially confirmed by a pediatrician, e.g., during a wellbeing check-up).
With the increasing number of training data, we may build more robust models that can yield more reliable screening results. 
Furthermore, with growing case numbers we could stratify model development, for example, by taking into account the age/weight/gender of the infant, or any other meta-information that is readily available.
Our current dataset is too small for such personalised modelling efforts.
However, with the ongoing data collection we will be in a position to explore such adaptation approaches.

\section{Conclusion}
\label{sec:lastwords}
Perinatal Strokes are difficult to detect in the first months of life, yet can have very serious, adverse consequences. 
Early detection of infants with Perinatal Stroke is essential in order to provide timely support and medical input to mitigate against adverse outcomes.
In this paper we have demonstrated that it is possible to use body-worn inertial measurement units and novel machine learning methods for sensor data analysis to automatically distinguish between the abnormal movements of a group of infants with Perinatal Stroke, and the normal movements of control infants. 
Through a rigorous evaluation of our method in a cohort of infants, who either had been diagnosed with Perinatal Stroke or were typically developing, we have laid the foundation for a screening tool that can potentially be used at population scale with minimal effort to enable early stage recognition of abnormal movements in young infants. 
Our method is straightforward to apply, inexpensive, and reliable with regards to the accuracy of analysis results.